\documentclass[pre,aps,twocolumn]{revtex4}
\begin{document}
\title{Solidity of viscous liquids. III. Alpha relaxation}
\author{Jeppe C. Dyre}
\affiliation{Department of Mathematics and Physics (IMFUFA), ``Glass and time'' - Danish National Research Foundation Centre for Viscous Liquid Dynamics, 
Roskilde University, Postbox 260, DK-4000 Roskilde, Denmark}
\date{\today}
\newcommand{\rok}{\rho_{\bf k}}
\newcommand{\xik}{\xi_{\bf k}}
\newcommand{\romk}{\rho_{-\bf k}}

\begin{abstract}
It is suggested that the $\omega^{-1/2}$ high-frequency decay of the alpha loss in highly viscous liquids, which appears to be generic, is a manifestation of a negative long-time tail as typically encountered in stochastic dynamics. The proposed mechanism requires that the coherent diffusion constant is much larger than estimated from the alpha relaxation time. This is justified by reference to the solidity of viscous liquids in an argument which, by utilizing the irrelevance of momentum conservation at high viscosity and introducing a center of mass diffusion constant, implies that at high viscosity the coherent diffusion constant is much larger than the incoherent diffusion constant.
\end{abstract}


\maketitle

\section{Introduction}

Viscous liquids and the glass transition have attracted considerable interest for many years \cite{kau48,har76,bra85,sch90,ang91,ang00}. The solidity of viscous liquids refers to their solidness, i.e., that a viscous liquid is similar to a disordered solids because molecular motion is mostly vibrational. This was the view, e.g., of Goldstein who in his now classic 1969 paper expressed the belief that ``when all is said and done, the existence of potential energy barriers large compared to thermal energy are intrinsic to the occurrence of the glassy state, and dominate flow, at least at low temperatures'' \cite{gol69}. This implies that solid-state elasticity arguments are relevant and useful for understanding viscous liquid dynamics \cite{gol69}. Actually, the view that viscous liquids are like solids by mostly being in elastic equilibrium is obvious already from Kauzmann's 1948 review, which referred to these large barrier transitions (flow events) as ``jumps of molecular units of flow between different positions of equilibrium in the liquid's quasicrystalline lattice'' \cite{kau48}. In the same spirit, Mooney in 1957 wrote that ``a liquid not only could be but perhaps should be treated as an elastic continuum with a stress relaxation mechanism'' \cite{moo57}.

Based on the solidity property a theory for the non-Arrhenius temperature dependence of the viscosity  was proposed some time ago, according to which the activation energy is proportional to the instantaneous shear modulus (the shoving model)  \cite{shov}. This was justified by arguing that during a barrier transition the surrounding liquid behaves like a solid subject to elastic deformation, and that most of the barrier energy is shear elastic energy located in the surroundings. In two subsequent papers (I and II, \cite{dyr99}) we further discussed the solidity of viscous liquids and introduced a ``solidity length,'' $l$, giving the length scale below which a viscous liquid is virtually indistinguishable from a disordered solid -- albeit one that flows. 

By proceeding inductively from experiment the present paper argues that solidity may explain generic features of the primary relaxation process in viscous liquids, the so-called alpha process. Alpha relaxation is the slowest and the dominant relaxation process. The alpha process is responsible for the glass transition, the kinetic falling out of equilibrium taking place when the cooling rate $d\ln T/dt$ exceeds the alpha relaxation rate \cite{kau48,har76,bra85,sch90,ang91,ang00}. Alpha relaxation is observable in dielectric relaxation experiments and in measurements of the frequency-dependent shear and bulk moduli \cite{har76,dex72} and, e.g., of the frequency-dependent specific heat \cite{chr85,bir85}. The loss peaks obtained by measuring these linear response properties are of similar shape, but usually not identical. The loss peak frequencies are also similar, but not identical, and their temperature dependencies are usually quite similar. In the following we shall not dwell on the dissimilarities. The focus is on the shear modulus, adopting the point of view that this is the primary response function while, e.g., the dielectric loss is determined by the coupling of molecular rotation to shear flow \cite{deb29,gem35,dim74,dia98,jak}.

The characteristic features of alpha response functions $\chi(\omega)$, as reported in the literature, are: 1) The loss peak frequency $\omega_m$ is non-Arrhenius with an activation energy which increases as temperature decreases; 2) on the low-frequency side of the loss peak the loss is virtually Debye like ($\propto\omega$); 3) on the high-frequency side there are significant deviations from Debye behavior, here the loss decays following an approximate power law: $\chi''(\omega)\propto\omega^{-n}$ ($n<1$).  In many cases a wing is observed a few decades above the loss peak frequency where $n$ changes to a lower value. 

The vast majority of experiments report values of $n$ between 0.3 and 0.7 \cite{boh93}. The exponent $n$ is often temperature dependent, implying a violation of time-temperature superposition. At the 1997 3rd International Discussion Meeting on Relaxations in Complex Systems it was suggested by Olsen that beta processes may play a role at much lower frequencies than generally expected, and that these could explain the wing \cite{ols98}. This was subsequently confirmed by Lunkenheimer and co-workers who, upon long-time annealing of propylene carbonate and glycerol, found that the wing indeed develops into a separate relaxation process \cite{lunk}. While the matter is still actively discussed, in the opinion of the Roskilde group the following picture emerges  \cite{ols01}: Once the effects of beta processes are minimized by going to sufficiently low temperatures (still in the equilibrium liquid phase), the generic features of alpha relaxation come to light. Generic alpha relaxation obeys time-temperature superposition, and for the high-frequency asymptotic behavior the generic alpha process is characterized by the universal exponent $n=1/2$:

\begin{equation}\label{1}
\chi_{\rm gen}''(\omega)\ \propto\ 
\left\{
\begin{array}{lr}
\omega & ,\ \omega\ll\omega_m\\
\omega^{-1/2} & ,\ \omega\gg\omega_m
\end{array}
\right.\,.
\end{equation}
Finally, generic alpha relaxation does not have universal loss peak widths. This conclusion was reached from dielectric relaxation measurements on several molecular liquids \cite{ols01}, but it applies also for the shear modulus $G(\omega)$ \cite{beh96,nis}, although shear mechanical measurements are considerably less accurate than dielectric measurements.

\section{The BEL model and its formal generalization}

Curiously enough, the conjecture that $n=1/2$ is a ``fixed point,'' which equilibrium viscous liquids approach at sufficiently low temperatures, represents a step backwards in time. Around 1970 it was noted that  $n$ is often close to $1/2$, and there was a considerable interest in explaining this theoretically. Some models, which over time have been suggested to explain Eq. (\ref{1}) or its equivalents, are: 
In 1948 Zener in his theory of ``relaxation by thermal diffusion'' for ultrasonic loss predicted that the internal friction at high frequencies varies as $\omega^{-1/2}$ \cite{zen48}. 
Glarum (in 1960), Doremus (1970), Phillips, Barlow, and Lamb (1972), Bordewijk (1975), Kimmich and Voigt (1978), Wyllie (1979), and Condat (1989) all arrived at the $\omega^{-1/2}$-tail from essentially one-dimensional diffusion models
\cite{gla60,dor70,phi72,bor75,kim78,wyl79,con89}. 
This is also basically the mathematical mechanism in the model of Isakovich and Chaban, who in 1966 predicted a limiting high-frequency sound absorption per wavelength in highly viscous liquids varying as $\omega^{-1/2}$ by regarding the liquid as a micro-inhomogeneous medium \cite{isa66} (see also Ref. \cite{lis97}).
In 1967 Barlow, Erginsav, and Lamb, in what is referred to as the BEL model, took an electrical engineer's approach and regarded the shear mechanical impedance of a viscous liquid as a parallel combination of a Newtonian liquid and a Hookean solid, implying an $\omega^{-1/2}$-tail for the high frequency shear modulus \cite{har76,bar67,lam78}.
Montrose and Litovitz in 1970 derived the $\omega^{-1/2}$-tail in a model invoking diffusion and relaxation of some unspecified order \cite{mon70}.
Majumdar, working in the time domain, arrived at the short time equivalent of the high-frequency tail in 1971 from the assumption that diffusive modes exist with amplitude $\propto k^{-2}$ \cite{maj71}.
Cunat in 1988 from a purely thermodynamic treatment arrived at a distribution of relaxation times on the logarithmic scale $\tau$ proportional to $\sqrt{\tau}$ for $\tau\rightarrow 0$, which implies the $\omega^{-1/2}$ high-frequency tail \cite{cun88}.
Sj{\"o}gren in 1990 showed that Eq.\ (\ref{1}) arises in mode-coupling theory under certain conditions \cite{sjo90}. 
Cichocki and Felderhof  \cite{cic94} in 1994 derived the $\omega^{-1/2}$ high-frequency tail in a three-dimensional diffusion model with a radial potential barrier, which is mathematically similar to the Isakovich-Chaban model.
-- Most of these models involve some species obeying the diffusion equation, which is assumed to apply also at times shorter than the alpha relaxation time \cite{dyr01}. This nontrivial assumption is also central for the ``long-time tail'' mechanism discussed below.

The BEL model focuses on the shear compliance $J(\omega)=1/G(\omega)$. In dimensionless units the BEL model has a definite prediction for $J(\omega)$, implying universal loss peak width. Because this prediction did not fit all experiments, the BEL expression was soon formally generalized by the same authors \cite{bar69} to

\begin{equation}\label{2}
J(\omega)\ =\ 
a + b\, (i\omega)^{-1/2} + c\, (i\omega)^{-1}\,.
\end{equation}
The original BEL model corresponds to $b^2=4ac$. Clearly, Eq.\ (\ref{2}) implies Eq.\ (\ref{1}) for the shear modulus loss $G''(\omega)$. In dimensionless units Eq.\ (\ref{2}), which we shall refer to as the ``generalized BEL expression,'' has one free parameter and thus nonuniversal loss peak width.

\section{The generalized BEL expression as a consequence of a long-time tail}

Equation (\ref{2}) gives a good representation of generic alpha relaxation. We proceed to use the fluctuation-dissipation (FD) theorem to translate Eq. (\ref{2}) into a statement concerning a velocity autocorrelation function. According to the FD theorem, if $Q$ is the $x$-direction displacement of the top of the liquid (area $L^2$ in the $x$-$y$-plane, height $h$ in the $z$-direction), the shear compliance is given by 

\begin{equation}\label{3}
J(\omega)\ =\ \ \frac{i\omega}{2\ k_BT}\,\frac{L^2}{h}\ 
\int_0^\infty\langle\Delta Q^2(t)\rangle\, e^{-i\omega t}\,dt\,.
\end{equation}
Here $\langle\Delta Q^2(t)\rangle$ is the equilibrium mean-square displacement of $Q$ in time $t$, i.e., the zero shear stress thermal average of the square of $\Delta Q(t)\equiv Q(t)-Q(0)$.

Via Eq.\ (\ref{3}) the generalized BEL expression Eq.\,(\ref{2}) translates into

\begin{equation}\label{4}
\langle\Delta Q^2(t)\rangle\ =\ 
A + B\, t^{1/2} + C\, t\,.
\end{equation}
In terms of the $Q$ mean-square displacement the velocity autocorrelation function is given by

\begin{equation}\label{5}
\langle\dot Q(0)\dot Q(t)\rangle \ =\ 
\frac{1}{2}\,\frac{d^2}{dt^2}\, \langle \Delta Q^2(t)\rangle\,.
\end{equation}
The generalized BEL expression is thus mathematically equivalent to

\begin{equation}\label{6}
\langle\dot Q(0)\dot Q(t)\rangle \ \propto\ -\ t^{-3/2}\,.
\end{equation}

Equation (\ref{6}), which is a simple way to summarize basic characteristics of generic alpha relaxation, immediately brings to mind the long-time tails discovered in 1967 by Alder and Wainwright in early molecular dynamics simulations of liquids \cite{ald67}. These authors showed that the velocity autocorrelation function for a single molecule does not decay exponentially to zero at long times as previously expected, but as a power law with exponent $-3/2$: 

\begin{equation}\label{7}
\langle \dot x(0)\dot x(t)\rangle\ \propto\ +\ t^{-3/2}\,\,\,(t\rightarrow\infty)\,. 
\end{equation}
The explanation is the following. According to statistical mechanics at any given time molecular velocities are uncorrelated. Consider a particular molecule at time $t=0$. It has a certain momentum while the surrounding molecules on average have zero momentum. As time passes the momentum of the molecule in question diffuses to the surrounding molecules (because the Navier-Stokes equation is basically a transverse momentum diffusion equation). At time $t$ the momentum has diffused the distance $\sim t^{1/2}$ away, thus spread among $\sim t^{3/2}$ molecules. If the initial momentum of the molecule in question is roughly equally shared among these, on average every molecule has momentum $\sim t^{-3/2}$. This includes the particular molecule in focus, thus explaining Eq.\ (\ref{7}). -- A nice classical review of long-time tails is that of Pomeau and Resibois from 1975 \cite{pom75}, while Kirkpatrick and co-workers recently gave an excellent review from a more general point of view \cite{kir02}.

Central to the above argument are {\it momentum conservation} and the {\it diffusion equation}. If Eq.\ (\ref{6}) were to reflect a long-time tail, two questions come to mind: 1) What is the role of momentum conservation in viscous liquids? 2) How does one explain a {\it negative} long-time tail \cite{hag97}?  

The first question arises because viscous liquids close to the calorimetric glass transition have viscosities roughly $10^{14}$ times larger than that of liquids like ambient water. Since the kinematic viscosity is the transverse momentum diffusion constant, these liquids have extremely large momentum diffusion constants. Consider a flow event taking the system from one minimum to another. If momentum conservation were relevant, a flow event could not result in movement of the center of mass. In reality the liquid is confined by container walls; these provide the external forces possibly needed to move the center of mass slightly to ensure that the molecules at the walls do not move. Thus just as momentum conservation plays no role for the description of point defect motion in crystals, momentum conservation is irrelevant for the description of viscous liquid dynamics under laboratory conditions \cite{note}.

Turning to the second question, it is shown in the next section that velocity autocorrelation functions are negative in any stochastic dynamics, i.e., dynamics described by a master equation.

What is the required conserved variable? Angular momentum conservation is irrelevant for the same reason that momentum conservation is. Energy conservation could be relevant because heat conduction is rather slow, but this process is almost temperature independent and thus seemingly unable to give alpha relaxation times, which depend dramatically on temperature. The only alternative left is {\it particle number conservation}.

Long-time tails arise from particle number conservation because density fluctuations at long times decay following the diffusion equation, leading to (where $\Delta\rho$ is the deviation from average density)

\begin{equation}\label{8}
\langle\Delta\rho({\bf r},0)\Delta\rho({\bf r}',t)\rangle\ 
\propto\ 
t^{-3/2}\exp\left(-\frac{({\bf r}-{\bf r}')^2}{4Dt}\right)\,.
\end{equation}
From Eq. (\ref{8}) it follows that any generic variable $F$ with zero average which is a [non-linear] function of density, decays following

\begin{equation}\label{9}
\langle F(0) F(t)\rangle \propto\ +\ t^{-3/2}\,\,\,(t\rightarrow\infty)\,.
\end{equation}
The Appendix gives a simple example of how Eq. (\ref{8}) implies Eq. (\ref{9}).

\section{Negative long-time tails in stochastic dynamics}

Viscous liquids are believed to be well described by stochastic dynamics \cite{gle98,sch00}. This could take the form of Langevin dynamics or, in the ultimate course-grained version, hopping between energy (``inherent dynamics'' \cite{sch00}). We proceed to show that the velocity autocorrelation function in any stochastic dynamics can be written as the negative of a state autocorrelation function. If $P_i$ is the probability for state $i$, the master equation is $\dot P_i=\sum_j\Lambda_{ij}P_j$. If the Green's function -- the probability to go from state $i$ to $k$ in time $t$ -- is denoted by $G_{i\rightarrow k}(t)$, and $P_{{\rm eq,}i}$ is the equilibrium probability of state $i$, Eq. (\ref{5}) implies

\begin{equation}\label{10}
\langle\dot Q(0)\dot Q(t)\rangle \ =\ 
\frac{1}{2}\sum_{i,j} (Q_i-Q_j)^2 P_{{\rm eq,}i}\,\frac{d^2}{dt^2}
G_{i\rightarrow j}(t)\,.
\end{equation}
Using the fact that the Green's function solves the master equation ($\dot G_{i\rightarrow j}=\sum_k\Lambda_{jk}G_{i\rightarrow k}$), detailed balance ($P_{{\rm eq,}i}G_{i\rightarrow j}(t)=P_{{\rm eq,}j}G_{j\rightarrow i}(t)$), and probability conservation ($\sum_i\Lambda_{ij}=0$), the right hand side of Eq. (\ref{10}) is reduced as follows:

\begin{eqnarray}\label{11}
\lefteqn{\frac{1}{2}\sum_{i,j,k} (Q_i-Q_j)^2 P_{{\rm eq,}i}\Lambda_{jk}\dot G_{i\rightarrow k}(t)}\nonumber\\
& \  = \ &
\frac{1}{2}\sum_{i,j,k} (Q_i-Q_j)^2 \Lambda_{jk} P_{{\rm eq,}k}
\dot G_{k\rightarrow i}(t)\nonumber\\
& \ =  \ & 
\frac{1}{2}\sum_{i,j,k,l} (Q^2_i+Q^2_j-2Q_iQ_j)  \Lambda_{jk} P_{{\rm eq,}k}
\Lambda_{il} G_{k\rightarrow l}(t)\nonumber\\
& \ = \ & 
-\sum_{i,j,k,l} (Q_i\Lambda_{il})(Q_j \Lambda_{jk}) P_{{\rm eq,}k}  G_{k\rightarrow l}(t)\,.
\end{eqnarray}
Thus in terms of the variable $B_k\equiv \sum_j Q_j\Lambda_{jk}$ we have \cite{note2}

\begin{equation}\label{12}
\langle\dot Q(0)\dot Q(t)\rangle \ =\ -\,\langle B(0) B(t)\rangle \,.
\end{equation}
Since autocorrelation functions of any state function are always positive in stochastic dynamics, this proves that the velocity autocorrelation function is always negative. Moreover, as argued above, generic autocorrelation functions decay as $t^{-3/2}$ for $t\rightarrow\infty$. We thus get the required Eq. (\ref{6}).

\section{Condition for long-time tails to operate at relatively short times}

The negative $t^{-3/2}$-decay applies as $t\rightarrow\infty$. The obvious question is how a long-time tail could play any role at times {\it shorter} than the alpha relaxation time. The only possibility, it seems,  is that density fluctuations, even at these relatively short times, decay follow the diffusion equation. Thus even for a few decades of time shorter than $\tau_\alpha$ the coherent intermediate scattering function $S(k,t)$ should obey

\begin{equation}\label{13}
\frac{S(k,t)}{S(k)}\ =\ \exp(-D k^2t)\,.
\end{equation}
If $a$ is the average intermolecular distance, the coherent diffusion constant defines a characteristic time $t_c$ via $t_c=a^2/D$. It is natural to expect that Eq. (13) applies only at times much longer than $t_c$. Thus the condition for the long-time tail mechanism to work above the alpha loss peak frequency is that $t_c\ll\tau_\alpha$, or 

\begin{equation}\label{14}
\frac{a^2}{\tau_\alpha}\ \ll\ D\,.
\end{equation}

The single-particle [incoherent] diffusion constant $D_s$ is defined by $D_s=\langle\Delta x^2(t)\rangle/2t$ ($t\rightarrow\infty$). One naively expects that information about the structure is basically forgotten when -- and only when -- the molecules on average have moved an intermolecular distance.  This would imply that $D_s\sim a^2/\tau_\alpha$, which may well be the case for some liquids. In several liquids, however, it has been observed that $D_s$ is much larger than $a^2/\tau_\alpha$ \cite{violations}. In either case, Eq. (\ref{14}) is clearly obeyed if the coherent and the incoherent diffusion constants decouple as follows:

\begin{equation}\label{15}
D_s\ \ll\ D\,.
\end{equation}
When asked what to expect if the coherent and incoherent diffusion constants decouple, a majority of researchers opted for the opposite inequality. Equation (\ref{15}) cannot be ruled out on experimental grounds, however, because there seems to be no measurements of the coherent diffusion constant for liquids just above the calorimetric glass transition.

\section{Diffusion constant decoupling as a consequence of solidity}

In this section we prove that the required diffusion constant decoupling is consistent with the solidity of viscous liquids. More accurately, it is shown that if the coherent diffusion constant is frequency independent, solidity implies diffusion constant decoupling. We consider $N$ molecules in volume $V$, however {\it not} with periodic boundary conditions. Density fluctuations are conveniently described in terms of the quantity $\rok$ defined \cite{boonyip} by the following sum over all molecular positions ${\bf r}_j$:

\begin{equation}\label{16}
\rok\ =\ \frac{1}{\sqrt N}\sum_j e^{i{\bf k\cdot r }_j}\,.
\end{equation}
The static structure factor is given by $S(k)=\langle\rok\romk\rangle$ and the coherent intermediate scattering function by  $S(k,t)=\langle\rok(t)\romk(0)\rangle$ \cite{boonyip}. In liquids and solids the long-wavelength limit of $S(k)$ is a very small number equal to the ratio between system compressibility and ideal gas compressibility at same density and temperature. Alternatively, the long-wavelength limit may be expressed in terms of the number $M$ of molecules in a large subvolume and its fluctuations: 
$S(k\rightarrow 0)=\langle\Delta M^2\rangle/\langle M\rangle$ \cite{boonyip,hanmc} (which is one for an ideal gas by Poisson statistics, but much smaller for liquids and solids). This number has two contributions, one from the phonon degrees of freedom and one from the inherent dynamics. The inherent part of this small number \cite{sti98} is thus even smaller. Henceforth $S(k)$ refers to only the inherent part of the structure factor, so we clearly have 

\begin{equation}\label{17}
S(k\rightarrow 0)\ \ll\ 1 \,.
\end{equation}

We shall assume that inherent density fluctuations are described by a linear Langevin equation,

\begin{equation}\label{18}
\dot\rok\ =\ -\gamma(k)\rok+\xik(t)\,.
\end{equation}
The last term is the usual Gaussian white noise characterized by $\xik^*=\xi_{-\bf k}$ and 
$\langle\xik(t)\xi_{{\bf k}'}(t')\rangle=2\gamma(k)S(k)\delta_{{\bf k}+{\bf k}',{\bf 0}}\delta(t-t')$, which ensures that $S(k)=\langle\rok\romk\rangle$. We proceed to argue that in the $k\rightarrow 0$ limit solidity implies that $\gamma(k)\propto k^2$ (as expected from particle conservation) and $D_s\ll D$, where $D$ is defined by $\gamma(k)=Dk^2$. In particular, Eq. (\ref{18}) implies Eq. (\ref{13}) for the intermediate scattering function. 

It is assumed that $V^{1/3}$ is smaller than the solidity length. This implies that in the time between two flow events there is elastic equilibrium throughout the sample \cite{dyr99}. We may thus adopt inherent dynamics \cite{sch00}, which regard flow events as instantaneous rearrangements taking the system from one energy minimum (inherent state \cite{sti83}) to another. 

To prove that solidity implies  $\gamma(k)\propto k^2$ as $k\rightarrow 0$ we first consider a single flow event. If $\Delta {\bf r}_j$ is the displacement of the j'th molecule, at small $\bf k$ the change in $\rok$ is given by 

\begin{equation}\label{19}
\delta\rok\ =\ \frac{1}{\sqrt N}\sum_j e^{i{\bf k\cdot r }\it _j}i{\bf k\cdot} \rm\Delta \bf r _{\it j}\,\,(k\rightarrow 0)\,.
\end{equation}
If $\bf r_{\rm 0}$ marks the location of largest molecular displacements, at small $\bf k$ the exponentials may all be replaced by $\exp(i{\bf k\cdot r }_{\rm 0})$. Thus, if $\bf R$ is the sum of all particle coordinates we get

\begin{equation}\label{20}
\delta\rok\ =\ \frac{e^{i{\bf k\cdot r }_{\rm 0}}} {\it\sqrt N} \,\it i {\bf k\cdot} \rm\Delta \bf R\,\,(k\rightarrow 0)\,.
\end{equation}

If momentum conservation were relevant, one would have $\rm\Delta \bf R{\rm  =}\bf 0$, because the center of mass is fixed. A characteristic feature of viscous liquids, however, is the already mentioned irrelevance of momentum conservation \cite{note}. Besides the coherent and single-particle diffusion constants there is consequently a third diffusion constant, which we shall refer to as the ``center of mass diffusion constant,'' $D_{\rm CM}$. In order to be well-defined in the $N\rightarrow\infty$ limit, writing ${\bf R}=(X,Y,Z)$ the center of mass diffusion constant should be defined by 

\begin{equation}\label{21}
D_{\rm CM}\ =\ \frac{\langle\Delta X^2(t)\rangle}{2Nt} \,\, (t\rightarrow\infty)\,.
\end{equation}
Here it is understood that the $N, V\rightarrow\infty$ limit is taken before $t\rightarrow\infty$. If  $D_{\rm CM}$ is frequency dependent, the above expression defines the dc limit. The high-frequency limit, $D_{\rm CM}(\infty)$, is determined by the mean-square displacement at short times:

\begin{equation}\label{22}
D_{\rm CM}(\infty)\ =\ \frac{\langle\Delta X^2(t)\rangle}{2Nt} \,\, (t\rightarrow 0)\,.
\end{equation}
Over a short time span flow events are uncorrelated.Thus, if the number of transitions per molecule per unit time is denoted by $\Gamma$ and $\langle\Delta X^2\rangle$ is the average squared change of X in one flow event, the change in $\rok$ in a short time $t$ is given by

\begin{equation}\label{23}
\langle |\Delta\rok(t)|^2\rangle\  =
\ N\Gamma t\,\frac{1}{N}k^2\langle \Delta X^2\rangle\ =\ 
\Gamma t k^2 \langle \Delta X^2\rangle\,\,(k\rightarrow 0)\,.
\end{equation}
On the other hand, the change in $\rok$ over a short time is determined by the noise term of Eq. (\ref{18}), leading to
$\langle |\Delta\rok(t)|^2\rangle=2\gamma(k)S(k)t$. Equating these two expressions leads to

\begin{equation}\label{24}
S(k\rightarrow 0)\ =\ 
\frac{\Gamma k^2 \langle \Delta X^2\rangle}{2\gamma(k\rightarrow 0)}\,.
\end{equation}
For the limit to be well defined and nonzero we must have $\gamma(k)\propto k^2$. 

Writing $\gamma(k)=Dk^2$, we proceed to show that solidity implies $D_s\ll D$. First note that, since $D_{\rm CM}(\infty)=\Gamma \langle\Delta X^2\rangle/2$, Eq. (\ref{24}) becomes

\begin{equation}\label{25}
S(k\rightarrow 0)\ =\ 
\frac{D_{\rm CM}(\infty)}{D}\,.
\end{equation}
From the inequality (\ref{17}) we conclude that

\begin{equation}\label{26}
D_{\rm CM}(\infty)\ \ll\ D\,.
\end{equation}
Consider a flow event taking the system from one minimum to another. In the time between two flow events there is elastic equilibrium. The liquid is indistinguishable from a disordered solids, and a flow event induces slight adjustments of the positions of molecules far from the relatively few molecules that move considerably \cite{sch00,dyr99}. If the liquid were completely homogeneous and isotropic before and after the flow event, the total displacement $\Delta \bf R$ would be zero. This is not realistic, however, so there is no reason to expect the total displacements to be zero. According to standard elasticity theory \cite{wyl79,dyr99,lan70,wyl80} displacements far from the center of the flow event are small, varying with distance as $r^{-2}$. Their contribution to $\Delta\bf R$ is insignificant, so we expect

\begin{equation}\label{27}
\sum_j\Delta{\bf r}_{\it j}^2\ 
\sim\ 
\Big( \sum_j\Delta{\bf r}_{\it j}\Big)^2\ \,.
\end{equation}
This implies that the high-frequency limits of $D_s$ and $D_{\rm CM}$ are comparable:

\begin{equation}\label{29}
D_s(\infty)\ \sim\ D_{\rm CM}(\infty)\,.
\end{equation}
Finally, we note that when diffusion constants are frequency dependent they always increase with frequency (see, e.g., Ref. \cite{dyr00}), and often quite a lot:

\begin{equation}\label{30}
D_s\ \ll\ D_s(\infty)\,.
\end{equation}

To summarize, by representing the solidity of viscous liquids by inherent dynamics we have argued that

\begin{equation}\label{31}
\frac{a^2}{\tau_\alpha}\ll D_s\ll D_s(\infty)\sim D_{\rm CM}(\infty)\ll D\,.
\end{equation}
The strong inequality sign $\ll$ was used liberally throughout this section (the inequalities all hold with $\ll$ replaced by $\le$, though): The first strong inequality presumably applies only for some liquids \cite{violations}, while for others there may be near equality. The second strong inequality is likely -- because this is what is generally observed for hopping in disordered solids \cite{dyr00} -- but not compelling. Only the third strong inequality is compelling, being a direct consequence of the small compressibility of liquids. Nevertheless, this is enough to establish the inequality (\ref{14}), once it is assumed that the coherent diffusion constant is frequency independent.

\section{Conclusion}

A long-time-tail mechanism \cite{oldref} provides a natural explanation for the seemingly generic $\omega^{-1/2}$ high-frequency decay of the alpha process. It has been argued that the required decoupling of coherent from incoherent diffusion follows from the solidity of viscous liquids. There are still problems to be addressed. One is that the derivation of the diffusion constant decoupling was based on the assumption that the sample is smaller than the solidity length. If the alpha relaxation time is one second, the solidity length is just a few thousand Angstroms \cite{dyr99}. The question arises how to deal with bulk viscous liquids. Given the fact that Eq. (\ref{18}) does not allow bulk volume relaxation on the alpha time scale, it is clear that something is missing \cite{note3}. Nevertheless, we surmise that the long-time tail mechanism survives the necessary supplements to arrive at a full theory. Another problem is that the constants $A$ and $C$ of Eq. (\ref{4}), which here appear as simple integration constants, cannot be arbitrary: In order for the generalized BEL expression to fit data these constants must be related to the constant $B$ by $AC\sim B^2$. A final challenge is to establish the consistency of the proposed scheme, where the coherent diffusion constant is assumed to be frequency independent whereas the incoherent diffusion constant most likely is strongly frequency dependent.

\appendix*\section{Long-time tails derived from the diffusion equation}
We here show an example of how long-time tails typically arise. By ``long time'' is meant times long enough that density fluctuations decay following the diffusion equation. As a simple case, consider a variable $F$ which is a sum of pairwise contributions, $F=\sum_{ij} \phi({\bf r}_i -{\bf r}_j)$. In terms of the density $\rho({\bf r})\equiv\sum_i\delta({\bf r}-{\bf r}_i)$ we have 

\begin{equation}\label{A1}
F\ =\ 
\int d{\bf r}d{\bf r}'\, \phi({\bf r}-{\bf r}')\,
\rho({\bf r})\rho({\bf r}')\,.
\end{equation}
If the deviation from average density is denoted by  $\Delta\rho\equiv\rho-\langle\rho\rangle$, $F$ is given as 

\begin{equation}\label{A2}
F\ =\ 
{\rm Const.}\,+\,
\int d{\bf r}d{\bf r}'\, \phi({\bf r}-{\bf r}')\,
\Delta\rho({\bf r})\Delta\rho({\bf r}')\,.
\end{equation}
The $F$-autocorrelation function is given by

\begin{eqnarray}\label{A3}
\langle F(0)F(t)\rangle & = & {\rm Const.}+\int d{\bf r}_1d{\bf r}_2d{\bf r}_3d{\bf r}_4\phi({\bf r}_1-{\bf r}_2)\phi({\bf r}_3-{\bf r}_4)\nonumber\\
\,&\times&\langle\Delta\rho({\bf r}_1,0)\Delta\rho({\bf r}_2,0)\Delta\rho({\bf r}_3,t)\Delta\rho({\bf r}_4,t)\rangle\nonumber\\
\,&\,&\,
\end{eqnarray}
In the Gaussian approximation the average of a product of four variables by Wick's theorem is a sum of three terms, each a product of two pair correlation functions. One of the three terms is time-independent and we get [assuming $\phi({\bf r})=\phi(-{\bf r})$] 

\begin{eqnarray}\label{A4}
\langle F(0)F(t)\rangle & = & {\rm Const.} +2\int d{\bf r}_1d{\bf r}_2d{\bf r}_3d{\bf r}_4\phi({\bf r}_1-{\bf r}_2)\phi({\bf r}_3-{\bf r}_4)\nonumber\\
\,&\times&\langle\Delta\rho({\bf r}_1,0)\Delta\rho({\bf r}_3,t)\rangle\langle\Delta\rho({\bf r}_2,0)\Delta\rho({\bf r}_4,t)\rangle\nonumber\\
\,&\,&\,
\end{eqnarray}
Letting $t\rightarrow\infty$ we find $\langle F\rangle^2={\rm const}$, so, if $\Delta F\equiv F-\langle F\rangle$, Eq.\ (\ref{A4}) implies 

\begin{eqnarray}\label{A5}
\langle \Delta F(0)\Delta F(t)\rangle & = & 2 \int d{\bf r}_1d{\bf r}_2d{\bf r}_3d{\bf r}_4 \phi({\bf r}_1-{\bf r}_2)\phi({\bf r}_3-{\bf r}_4)\nonumber\\ 
\,&\times&\langle\Delta\rho({\bf r}_1,0)\Delta\rho({\bf r}_3,t)\rangle\langle\Delta\rho({\bf r}_2,0)\Delta\rho({\bf r}_4,t)\rangle\nonumber\\
\,&\,&\,
\end{eqnarray}
A standard assumption in liquid theory is that the density correlation function $\langle\Delta\rho({\bf r},0)\Delta\rho({\bf r}',t)\rangle$ regarded as a function of 
${\bf r}'$ and $t$ obeys the diffusion equation at long times. Since the density correlation function goes to zero as $|{\bf r}-{\bf r}'|\rightarrow\infty$ and 
$t\rightarrow\infty$, we have (where $D$ is the diffusion constant)

\begin{equation}\label{A6}
\langle\Delta\rho({\bf r},0)\Delta\rho({\bf r}',t)\rangle\ 
\propto\ 
t^{-3/2}\exp\left(-\frac{({\bf r}-{\bf r}')^2}{4Dt}\right)\,.
\end{equation}
When this is substituted into Eq.\ (\ref{A5}), one finds at long times, if $\phi$ is reasonably short ranged, $I\equiv\int d{\bf r}\phi({\bf r})$, and $V$ is the volume:

\begin{equation}\label{A7}
\langle \Delta F(0)\Delta F(t)\rangle \propto\ +\ V\ I^2\
t^{-3/2}\,.
\end{equation}


\begin{thebibliography}{99}
\bibitem{note} As an example, consider a viscous liquid with alpha relaxation time equal to one second. In this case momentum conservation is relevant for measurements on second time scales only if the sample has linear dimensions at least in the kilometer range.
\end{thebibliography}
\end{document}